\documentclass[%
preprint,
 amsmath,amssymb,
 aps,
]{revtex4-2}
\usepackage{xcolor}
\usepackage{graphicx}
\usepackage{dcolumn}
\usepackage{bm}

\begin{document}

\title{Magnetic Signature of Thermo-Electric Cardiac Dynamics}

\author{Anna Crispino\textsuperscript{1,*}, Martina Nicoletti\textsuperscript{1,2,*}, Alessandro Loppini\textsuperscript{3,°°}, Alessio Gizzi\textsuperscript{2}, Letizia Chiodo\textsuperscript{1}, Christian Cherubini\textsuperscript{4}, Simonetta Filippi\textsuperscript{1,5}}

\affiliation{\textsuperscript{1}Department of Engineering, Università Campus Bio-Medico di Roma, Italy
}%

\affiliation{ \textsuperscript{2}Center for Life Nano and Neuro Science (CLN2S@Sapienza), Italian Institute of Techlology, Rome, Italy
}%

\affiliation{\textsuperscript{3}Department of Medicine and Surgery, Università Campus Bio-Medico di Roma, Italy
}%
\affiliation{\textsuperscript{4}Department of Science and Technology for Sustainable Development and One Health, Università Campus Bio-Medico di Roma, Italy
}%
\affiliation{\textsuperscript{5}Istituto Nazionale di Ottica del Consiglio Nazionale delle Ricerche (CNR-INO), Florence, Italy}

\affiliation{\textsuperscript{*}These authors contributed equally to this work}
\affiliation{\textsuperscript{**}Corresponding author: a.loppini@unicampus.it}

\begin{abstract}

Developing new methods for predicting electromagnetic instabilities in cardiac activity is of primary importance. However, we still need a comprehensive view of the heart's magnetic activity at the tissue scale. To fill this gap, we present a model of soft active matter, including thermo-electric coupling, suitably modified to reproduce cardiac magnetic field. Our theoretical framework shows that periodic stimulations of cardiac cells create an external magnetic field evidencing restitution features of nonlinear cardiac dynamics and magnetic restitution curves better discriminate instabilities and bifurcations in cardiac activity. This new framework lays the foundation for innovative, non-invasive diagnostic tools for cardiac arrhythmias.
\end{abstract}

\maketitle

\section{Introduction}
Over the past fifty years, the adoption of preventive strategies targeting cardiac risk reduction has significantly decreased mortality from cardiovascular causes \cite{Stewart2020, Rippe2018}. Despite this considerable progress, cardiovascular diseases remain the leading cause of death in developed countries, with sudden cardiac death (SCD) constituting approximately half of all cardiac-related fatalities \cite{Zheng2001, DiCesare2024}. In this scenario, understanding the basic mechanisms for identifying risk factors and therapeutic approaches is critical for diagnosing, preventing, and managing cardiac arrhythmias. Electric instabilities in cardiac dynamics, arising in the form of action potential alternans in time and space \cite{Gizzi2013}, are well-established precursors of arrhythmias related to SCD \cite{Qu2014,Karma2013}. Spatiotemporal oscillations are attributed to multiscale and multiphysics factors, starting from cellular-level ion dynamics, comprising inherent tissue heterogeneity and anisotropy and entailing external factors like thermo-mechanical variation~\cite{kiyosue1993ionic, crozier1926distribution, FEDOROV20081587, Filippitemperature, Gizzi2017, correlation_temperature_2021, Crispino2024cross}. Increased body temperature occurs in the case of fever, sports fatigue, heat strokes, and in the medical practice, in the case of whole-body hyperthermia-induced to improve the chemotherapy success rate. Low body temperatures are used in medical practice to minimize cellular damage. Specifically, thermal monitoring is widely used as a therapeutic intervention to prevent injuries associated with cardiac arrest \cite{MOORE2011}. Body temperature can also decrease dramatically during accidents as avalanches and ship sinkings for instance. From a cellular point of view, changes in body temperature result in oscillation in action potential duration (APD), which is considerably increased at low temperatures (Figure~\ref{fig:fig_0}b). In cardiac electrophysiology, a characteristic curve known as the restitution curve can be defined (Figure~\ref{fig:fig_0}a) \cite{Nolasco_restitution,Cherry_2004} to describe how the APD depends on the recovery phase between each cardiac cycle, i.e., the diastolic interval (DI), or, alternatively, on the stimulation period, i.e., the pacing cycle length (PCL) (Fig. S1, Supplementary material). 
\begin{figure}[ht]
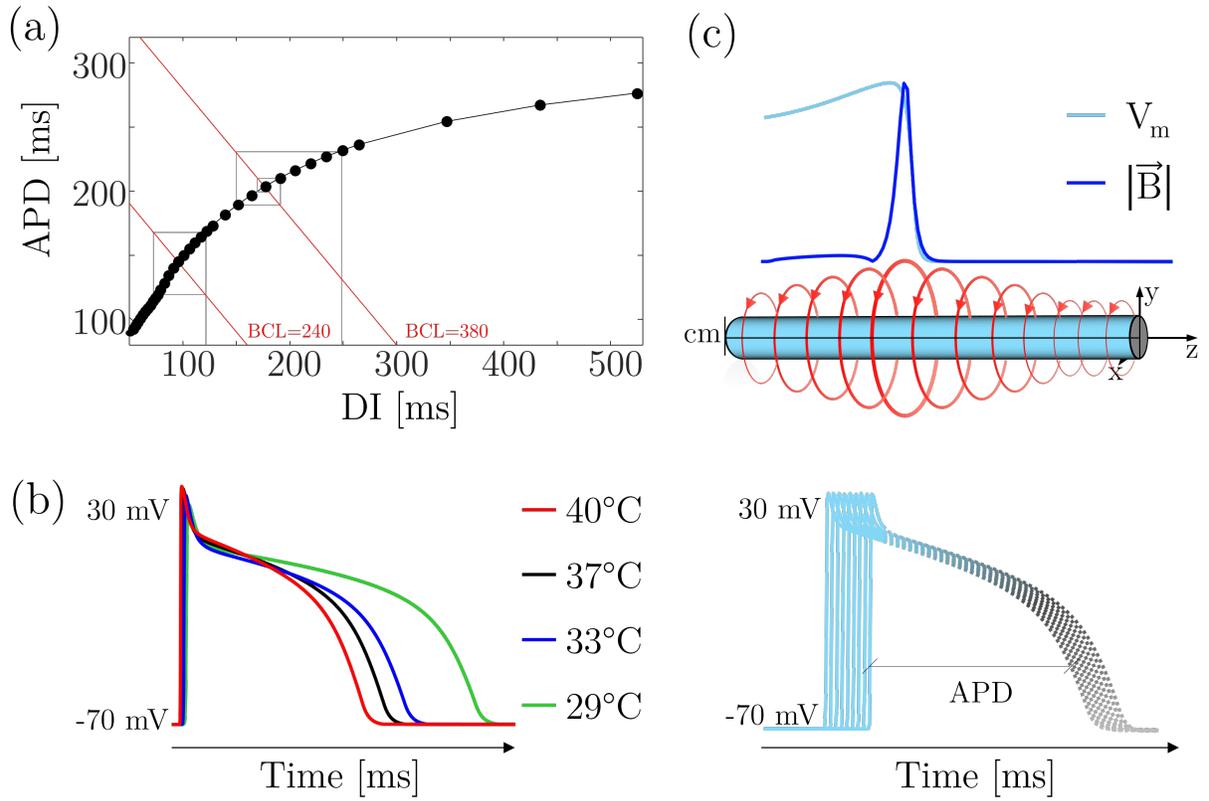

    \caption{\small Schematic representation of a (a) APD-DI iterated restitution map with cobweb examples, (b) thermo-electric morphology of AP over time, and (c) 1D cylindrical geometry model supporting cardiac AP spatiotemporal propagation and magnetic field calculation.}    
\label{fig:fig_0} 
\end{figure}

The analysis of restitution curves provides valuable insights into the dynamical behavior of cardiac electrical activity, describing the transition between normal and arrhythmogenic regimes. These regimes are distinguished in the restitution curves by changes in the slope $s$. Steep regions, with $|s|>1$, are, indeed, associated to beat-to-beat oscillation, i.e. alternans, leading to conduction block and spiral wave breakup \cite{Karma_spiral_breakup}. On the other hand, in smooth regions ($|s|<1$) the beat rhythm is normal.

The coordinated electrical activity resulting from the AP propagation gives rise to a magnetic field (Figure~\ref{fig:fig_0}c). A fundamental question in the study of cardiac magnetic fields is whether they convey more physical information than the respective electric field, which is investigated with standard and well-established techniques. In the past sixty years, both experimental and theoretical works demonstrated the capability of the magnetic field to reveal information not found in the electric signature of cardiac excitation~\cite{ROTH1986, ROTH1988, plonsey_nature_1982, Holtzer2004}, paving the way to the establishment of magnetocardiography as an advanced technique for the diagnosis of cardiac pathologies~\cite{roth2023biomagnetism,roth2024magnetocardiogram}. An open challenge in the study of cardiac magnetic field is the investigation of magnetic signals at the cellular scale, because of their very low intensity (in pT-nT range)~\cite{barry2016optical,webb2021detection,baudenbacher2002high,tanaka2003measurement,arai2022millimetre} which poses several experimental challenges.
Emerging technologies, such as bio-compatible quantum sensing~\cite{barry2016optical,webb2021detection,arai2022millimetre}, hold the promise to shed light on the molecular origin of biological magnetic fields, and, at the same time, to further enhance diagnostic accuracy, potentially transforming the landscape of arrhythmia management in the near future.
In this intriguing scenario, mathematical models could significantly contribute in expanding our comprehension of the cardiac function by offering an organic description of the complex interplay of electric and magnetic dynamics.

In this Letter, we unveil the magnetic features of an excitable biological system by considering a spatio-temporal temperature-dependent cardiac scenario. To introduce such a perspective, we deploy innovative magnetic field-based indicators by selecting a minimal case study represented by a one-dimensional (1D) geometry. An extended validation of predictability is obtained by comparing and contrasting with standard action potential (AP)-based indicators, such as duration (APD) and conduction velocity (CV), considering a wide range of temperatures, thus contributing to the definition of a novel thermo-electro-magnetic multiphysics theoretical framework.

\section{Methods} 

Cardiac AP excitation was obtained using a four-variable phenomenological model, fine-tuned on optical mapping experimental recordings~\cite{Gizzi2013}, including temperature-related effects~\cite{fenton2013role}, and amenable for full pacing-down restitution protocol (see Fig.~\ref{fig:fig_0}).
\begin{eqnarray}
\label{eq1}
    \partial_t u&=& D\nabla^2u-(J_{fi}+J_{si}+J_{so}) \\
    \label{eq2}\nonumber
    \partial_t v&=&\Phi_v(T)\left[(1-H(u-\theta_v)) \frac{v_\infty -v}{\tau_v^-}-\frac{H(u-\theta_v)}{\tau_v^+} \right]\\
    \label{eq3}\nonumber
    \partial_t w&=&\Phi_w(T)\left[(1-H(u-\theta_w)) \frac{w_\infty -w}{\tau_w^-}-\frac{H(u-\theta_w)}{\tau_w^+} \right]\\
    \label{eq4}\nonumber
    \partial_t s&=& \Phi_s(T) \left[ \frac{(1+\tanh(k_s(u-u_s)))/2-s}{\tau_s} \right]
\end{eqnarray}
Here, the variable \textit{u} represents the non dimensional membrane potential, while \textit{v}, \textit{w}, and \textit{s} regulate temperature-dependent fast inward ($J_{fi}$), slow inward ($J_{si}$), and slow outward ($J_{so}$) currents (see Supplementary Material for details).
Current density $\vec J$ resulting from action potential propagation was estimated using Ohm's law, considering for the sake of simplicity homogeneous and isotropic tissue conductivity $\sigma$ and a dimensional membrane potential varying in the z coordinate only $V_{m}(t,z)$ (with current density flowing on the z direction only then): \cite{fenton1998vortex,fenton2013role}:
\begin{equation}
\label{eq19}
     \vec J=-\sigma \nabla V_m\longrightarrow J_z(t,z')=-\sigma \frac{\partial V_m}{\partial z}
\end{equation}
For further details on tissue conductivity and dimensional membrane potential estimation we refer the reader to the Supplementary material.
We need to approach the magnetic problem in the simplest possible way while still grasping the relevant physics. The extremely slow variations of the electric (and consequently of the magnetic) field involved compared to signals moving at the speed of light allow us to neglect any electromagnetic wave phenomena and safely use magneto-static formulas of classical electromagnetism adapted to the present framework. Having in mind a straight cylinder in the z-direction with $z'\in[z_{min},z_{max}]$ (Fig.~\ref{fig:fig_0}), with an orthogonal circular section of radius $R'$ in the $x'y'$ plane, denoted by $S'=\pi R'^2$ and current flowing along the z direction only i.e. $\vec J\equiv(0,0,J_z)$, we can adopt standard Electromagnetism tools in order to obtain the magnetic field $\vec B_0$ around it in vacuum (or equivalently in the air). As usual in the Literature, primed quantities refer to the specific source portion location and not primed ones to the position in which the magnetic field is computed, while the corrensponding primed and non primed Cartesian coordinate systems can be exactly superimposed. 

To cope with minimal mathematical complexity, the section is considered extremely small so that we can use the assumption of dealing with a 1 D straight line current source along the z-axis such that $J_z$ is constant along the section orthogonal to the z-direction and so $I(t,z')=\int_{S'}\vec J(t,\vec r')\cdot d\vec S'= 
J_z(t,z')\pi R'^2$. Using  Biot-Savart's law for a zero thickness wire
\begin{equation}
\vec B_0=\frac{\mu_0}{4\pi}\int_{z_{min}}^{z_{max}}\frac{I(t,z')d\vec l' \wedge \Delta \vec r}{\vert \Delta \vec r\vert^3}
\label{eq3}
\end{equation}
with $\Delta \vec r=\vec r -\vec r'$, $\vec r =(x,y,z)$, $\vec r' =(0,0,z')$ and $\vec dl' =(0,0,dz')$,
simple vector analysis ($\hat i$ is the versor along the x axis and $\hat j$ along the y one) leads to:
\begin{eqnarray}\vec B_0&=&-\frac{\mu_0}{4\pi}\int_{z_{min}}^{z_{max}}\frac{I(t,z')y\,dz'}{[x^2+y^2+(z-z')^2]^{\frac32}}\hat i+\nonumber\\ &+&\frac{\mu_0}{4\pi}\int_{z_{min}}^{z_{max}}\frac{I(t,z')x\,dz'}{[x^2+y^2+(z-z')^2]^{\frac32}}\hat j\,.\label{eq4}\end{eqnarray}
If $ z_{min}\to-\infty$, $ z_{max}\to + \infty$ and $I(t,z')=I_0$ with $I_0$ being constant, the above formula leads to
\begin{equation} 
\vec B_0=-\frac{\mu_0 I_0 y}{2\pi(x^2+y^2)}\hat i+ \frac{\mu_0 I_0 x}{2\pi(x^2+y^2)}\hat j
\label{eq5}
\end{equation}  
which is the Cartesian coordinate version of the field generated by a thin indefinite straight wire carrying a constant electrical current, which in cylindrical coordinates is well known to be $\vec B_0=\frac{\mu_0 I}{2\pi r}\hat \phi$. 
If instead we would have worked in a situation in which the section of the wire would have not been negligible, we should have adopted the well known electromagnetism formula: 
\begin{equation}
\vec B_0=\frac{\mu_0}{4\pi}\int_{\tau'}d^3x'\frac{\vec J(t,\vec r')\wedge \Delta \vec r}{\vert \Delta \vec r\vert^3}
\label{eq6}
\end{equation}
requiring more involved multidimensional integrals on the volume $\tau'$ of the wire. Incidentally, expanding in Taylor series the above formula on the point $\vec{r}'=(0,0,z')$, keeping the zero order term, and integrating $dx'dy'$ on the orthogonal sections $S'$, one easily recovers Eq.~\ref{eq4}.

Magnetic restitution values were extrapolated for each PCL by determining the peak value across the complete temporal span of each beat (n-1 and n beats) and the entire length of the wire, within a 200 $\mu$m radius from the wire's center (Figure~\ref{fig:fig_1}d). 

For the APD\textsubscript{80} restitution curves, the action potential duration was evaluated by thresolding simulated signals at 80\% of repolarization and averaging over the entire wire length for each PCL, both for the \textit{n-1} and \textit{n} beats datasets.

\subsection{RESULTS}

Traditional indicators based on AP features, specifically APD and CV, were compared with novel indicators derived from the magnetic field (Figure~\ref{fig:fig_1}). As expected APD increases at lower temperatures, while CV decreases (Figure~\ref{fig:fig_1}a, b) \cite{Gizzi2017, Crispino2024cross, fenton2013role}.

\begin{figure*}[ht!]
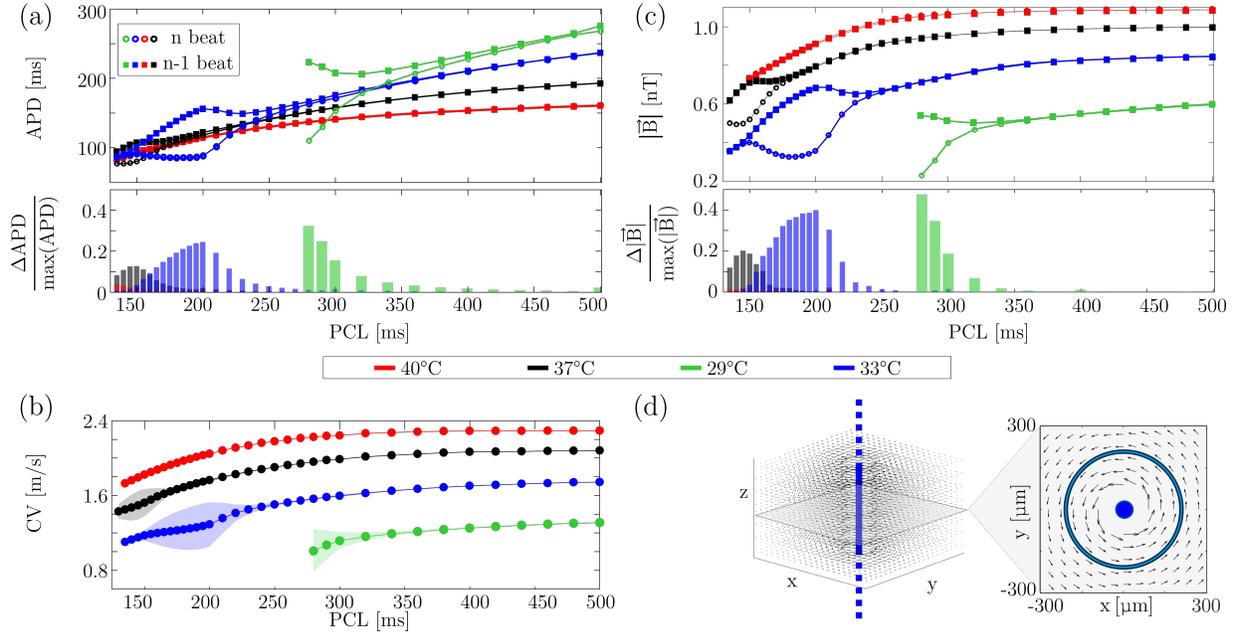

  \caption{\small Restitution curves of (a) APD and (b) CV (mean $\pm$ standard deviation), and (c) magnetic field norm at four different temperatures (40$^\circ$C to 29$^\circ$C). Panel (d) shows the superimposition at different time frames of the three-dimensional  magnetic field configuration at $37^\circ$~C, including the 2D representation in the \textit{x-y} plane for a selected \textit{z} coordinate. The blue annular ring ($r = 200~\mu$m) represents the chosen spatial grid for computing the magnetic field norm.}    
\label{fig:fig_1} 
\end{figure*}

In terms of the corresponding magnetic field, we confirm a peak values between 0.1~pT-1~nT in line with the literature~\cite{barry2016optical,arai2022millimetre,tanaka2003measurement,baudenbacher2002high}. This value decreases as temperature decreases, exhibiting an opposite trend with respect to APD (Figure~\ref{fig:fig_1}c). Such an effect is due to the slowing down of AP propagation and to the smoothing of the AP depolarization and repolarization wavefronts, actually reducing the intensity of magnetic field sources at the cellular level, i.e., diffusive currents.

In relation to beat-to-beat variations of cardiac activation maps, both the electrical-based indicators and magnetic field capture alternans regimes, whose onset is quantified by bifurcation points on restitution curves. In this regard, we compared the propensity of the indicators in highlighting alternans by computing the difference of their normalized value between consecutive beats at different temperatures.
It is well-established that hypothermic conditions promote the onset of alternans, which is indicated by the earlier bifurcation at lower temperatures. This effect is even more pronounced in the magnetic field curves, where the maximum beat-to-beat difference between normalized values is 0.5, compared to 0.3 for the APD for the lowest temperature. This suggests that magnetic field indicators are more sensitive in detecting the onset of these complex spatiotemporal dynamics under varying thermal conditions.

In this regard, to gain a comprehensive understanding of alternans through the perspective of the magnetic field, we characterized its spatial distribution by conducting a cross-temperature analysis of the magnetic energy density ($u_{m}=B^2/(2\mu_0)$) in space. For this analysis, we selected at each temperature three PCLs representative of non-alternating regime, alternans onset, and maximum alternans amplitude (Fig.~\ref{fig:fig2}), and computed the  magnetic energy density in the space surrounding the wire. 

\begin{figure*}[ht!]
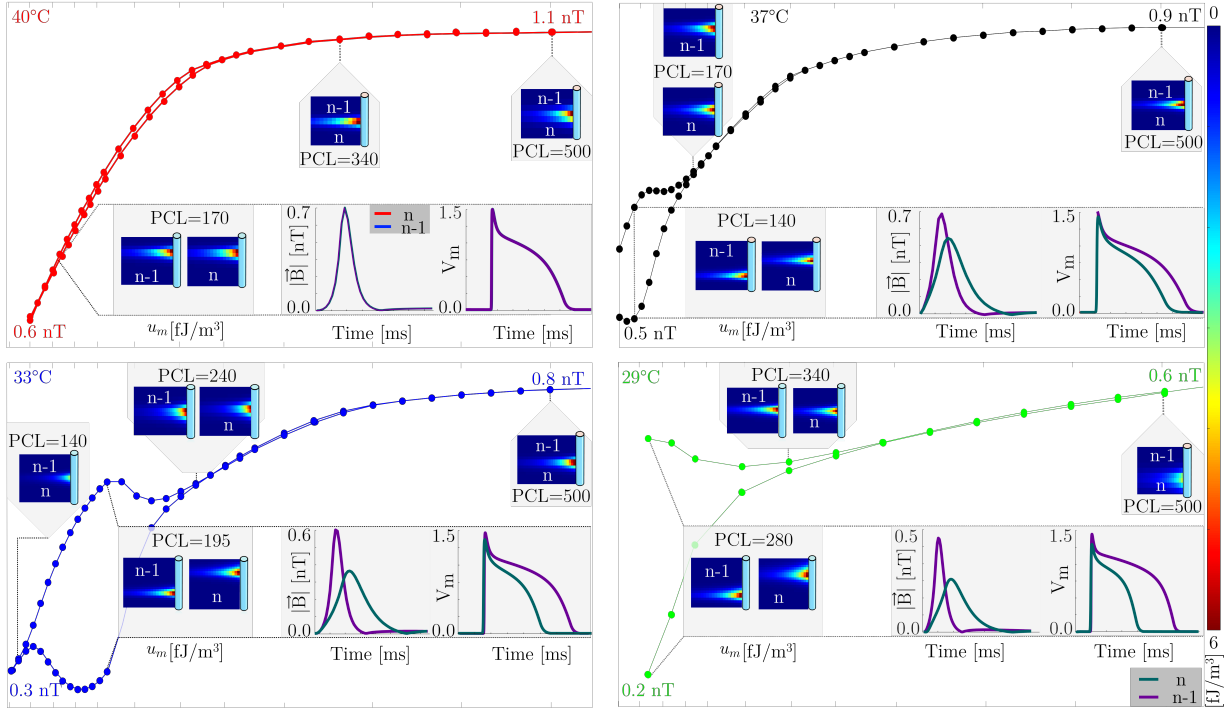

    \caption{\small Representation of the magnetic energy distribution in the  magnetic restitution curves. For each restitution curve we show the distribution along the wire of the magnetic energy density ($u_m$ [fJ/$m^3$]) for three selected PCLs, representative of steady-state, alternans onset and maximum alternans amplitude. Also, for the maximal alternans amplitude we show the action potentials and magnetic field norm for the $n$ (in  green) and $n-1$ (in purple) beat.}
    \label{fig:fig2}
\end{figure*}

In the case of maximum alternans amplitude, we also examined the electrophysiological and magnetic field temporal responses between two consecutive beats. At all the selected PCLs the magnetic energy distribution mirrored the spatial distribution of the magnetic field, showing an increased density in the proximity of the active region where the AP wavefront is localized (Figs.~\ref{fig:fig_0}~(c) and~~\ref{fig:fig2}). It is worth to note that the magnetic energy distribution is strongly related to the AP features (Fig.~\ref{fig:fig2}). Specifically, short APs (\textit{n-1} beat) produce a broader temporal and spatial distributions of the magnetic field compared to long AP (\textit{n} beat). In addition, for short APs we observed a notable decrease in magnetic field intensity and in the corresponding energy density in the space. This extended temporal and spatial magnetic activity associated with short APs might be related to a slow depolarization phase, which is in turn temperature-dependent (Fig.~S1, Supplementary material) \cite{ROTH1986}. Indeed, the velocity of the upstroke (represented by $dV_m/dt_{MAX}$, Fig.~S1 (b)) significantly affects the spatial and temporal gradients of membrane voltage, and thus the current density generating the magnetic field. Also, the decreased intensity of the magnetic field mainly reflects reduced diffusive currents as sources of magnetic activity. 
Moreover, our findings indicate that as the temperature decreases, the magnetic energy density appears to be more distributed along the propagation direction and its surrounding regions. This suggests a broader spatial distribution of magnetic activity under hypothermic conditions, which may further affect the dynamics of cardiac excitation and conduction.

The analysis of spatial and temporal distribution both of the magnetic field intensity $\vert \vec B \vert$ and of magnetic energy density in a simple geometry, as the straight wire considered here, could be important for understanding and optimizing the performance of highly sensitive devices such those based on nitrogen-vacancy (NV) centers, a particular class of magnetometers well known for their bio-compatibility and ability to operate at room temperature. The measure of magnetic fields via NV-centers relies on Zeeman effect~\cite{barry2020sensitivity,moreva2020practical,schirhagl2014nitrogen}. Consequently, the higher the intensity of B, the more intense is the resulting coupling between the magnetic field and spin of the quantum sensor, and therefore the observed splitting in the electron spin resonant spectra.

\section{DISCUSSION AND CONCLUSION}
Gaining a deeper comprehension of the origin and progression of cardiac arrhythmias is of critical importance for the development of novel diagnostic and therapeutic strategies. In this context, the study of cardiac magnetic field could offer an interesting perspective, that already revealed its potential at diagnostic level~\cite{smith2006comparison}. However, a magnetic description of cardiac alternans at the cellular level is still missing, in particular as far as temperature-related effects are considered. In this Letter, we aim to bridge the gap between traditional electrophysiological indicators and novel magnetic field measurements from a cellular scale point of view.

We characterized in a simple system, a straight wire, cardiac alternans by comparing tradition APD-PCL restitution curves with novel restitution curves based on the magnetic field produced by simulated cardiac electrical activity at different temperatures. Our simulations are in agreement with existing studies highlighting an augmented risk of cardiac alternans at low temperatures~\cite{kiyosue1993ionic, crozier1926distribution, FEDOROV20081587, Filippitemperature, Gizzi2017, correlation_temperature_2021, Crispino2024cross}, where the traditional APD indicator has the least good performance in the identification of the alternans onset (Fig.~\ref{fig:fig_1}). In contrast, the magnetic field restitution curves clearly highlight the onset of alternans and beat-to-beat differences are enhanced with respect to the APD restitution curves (Fig.~\ref{fig:fig_1}). Overall, our results indicate that magnetic field based indicators could substitute and even outperform traditional electrical indicators in detecting cardiac alternans especially at low temperatures (Fig.~\ref{fig:fig_1}).

Also, we show that action potential features are strongly related to the corresponding temporal and spatial behaviour of the magnetic field. In particular, the intensity and the spatial distribution of the resulting field mainly determined by the upstroke phase of the cardiac AP, which is also influenced by temperature. This strong relation between the magnetic field and cellular electrical activity offers interesting possibilities for the non-invasive investigation of cardiac electrophysiology in both healthy and diseased cases, allowing the study of biological samples in physiological conditions, even without a direct access to the cells. 

It is important to underline that this study is performed on a simplified geometry, which allows the characterisation of alternans in the temporal regime. However, it is well-known that cardiac alternans have also complex spatial behaviors~\cite{Crispino2024cross,hayashi2007dynamic,de2008spatially}. Therefore, future developments of this work will specifically focus on the spatial characterization of magnetic field also taking into account of the tissue heterogeneity ans anisotropy~\cite{kiyosue1993ionic, crozier1926distribution, FEDOROV20081587, Filippitemperature, Gizzi2017, correlation_temperature_2021, Crispino2024cross}, which is essential to originate complex spatial and temporal dynamics.

Overall, this study constitutes a starting point not only for a further theoretical investigation of the cardiac magnetic field in more complex configurations, e.g. in cell cultures or tissue patches, but also could help to shed light on molecular mechanisms at the basis of cardiac arrhythmias, especially in the case of disturbances associated to rare genetic disorders. Moreover, such a theoretical framework could serve as a guide for the development of new experimental tools for the detection of cellular magnetic fields, and help in the interpretation of the experimental data taking into account for the biophysical mechanisms at the basis of cardiac action potentials. 

In conclusion, our findings highlight the potential of magnetic field indicators to provide a more accurate and comprehensive diagnostic tool for cardiac electrophysiology, paving the way for future advances in non-contact cardiac diagnostics.

\subsection*{Funding}
This research has been funded by the European Commission-EU under the HORIZON Research and Innovation Action MUQUABIS GA n. 101070546, and by the European Union - NextGeneration EU, within PRIN 2022, PNRR M4C2, Project QUASAR 20225HYM8N [CUP C53D2300140 0008].\\

\begin{acknowledgments}
We wish to acknowledge the support of the Italian National Group for Mathematical Physics, GNFM-INdAM.
\end{acknowledgments}

\bibliography{biblio}
\newpage
\begin{figure}[ht]
\setcounter{figure}{0}    
    \includegraphics[width=\columnwidth]{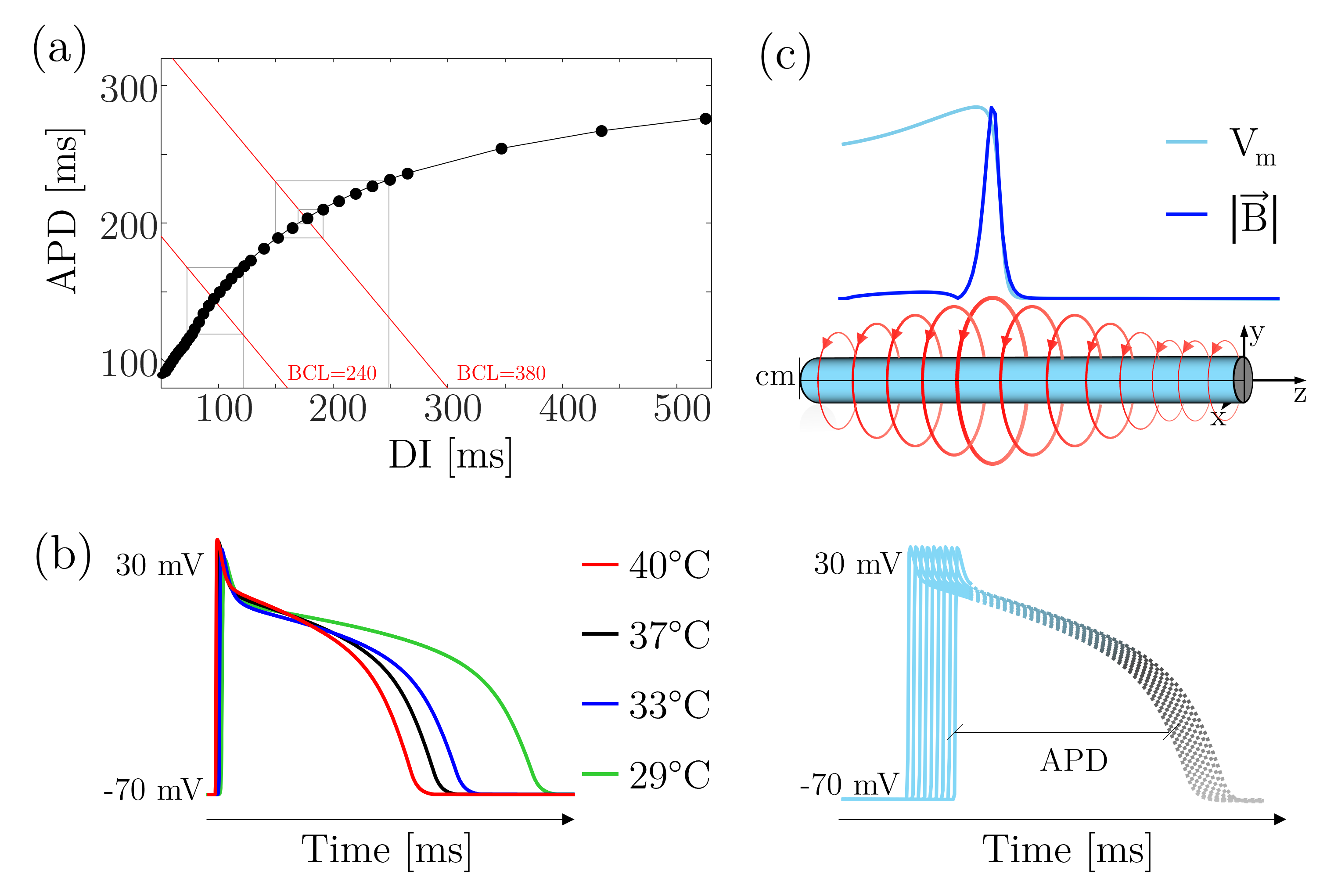}
   \caption{\small     Schematic representation of a (a) APD-DI iterated restitution map with cobweb examples, (b) thermo-electric morphology of AP over time, and (c) 1D cylindrical geometry model supporting cardiac AP spatiotemporal propagation and magnetic field calculation.}    
\label{fig:fig_0} 
\end{figure}

\newpage
\begin{figure*}[h]
    \includegraphics[width=\textwidth]{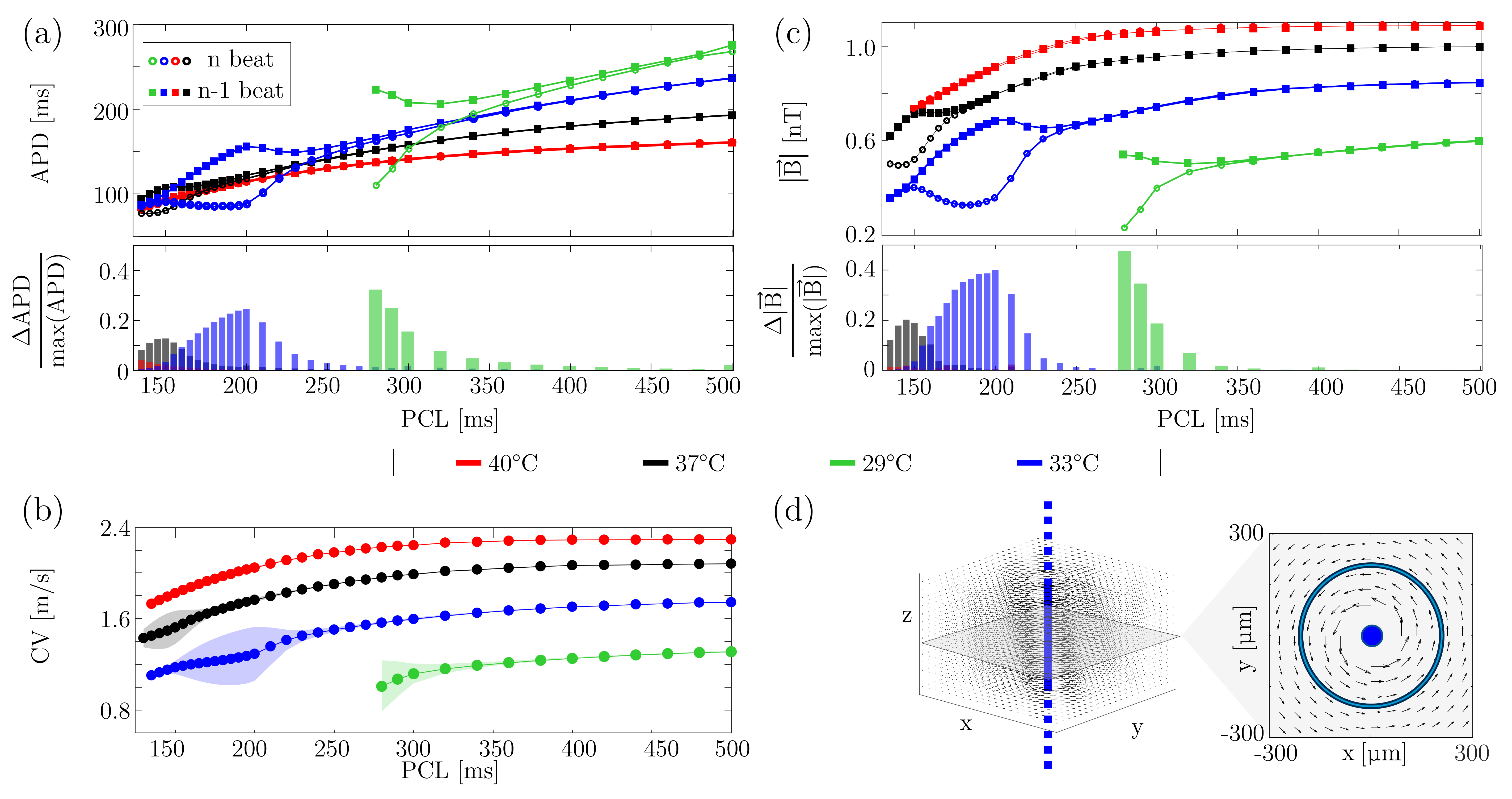}   
    \caption{\small Restitution curves of (a) APD and (b) CV (mean $\pm$ standard deviation), and (c) magnetic field norm at four different temperatures (40$^\circ$C to 29$^\circ$C). Panel (d) shows the superimposition at different time frames of the three-dimensional  magnetic field configuration at $37^\circ$~C, including the 2D representation in the \textit{x-y} plane for a selected \textit{z} coordinate. The blue annular ring ($r = 200~\mu$m) represents the chosen spatial grid for computing the magnetic field norm.}    
\label{fig:fig_1} 
\end{figure*}

\newpage

\begin{figure*}[h]
    \includegraphics[width=\textwidth]{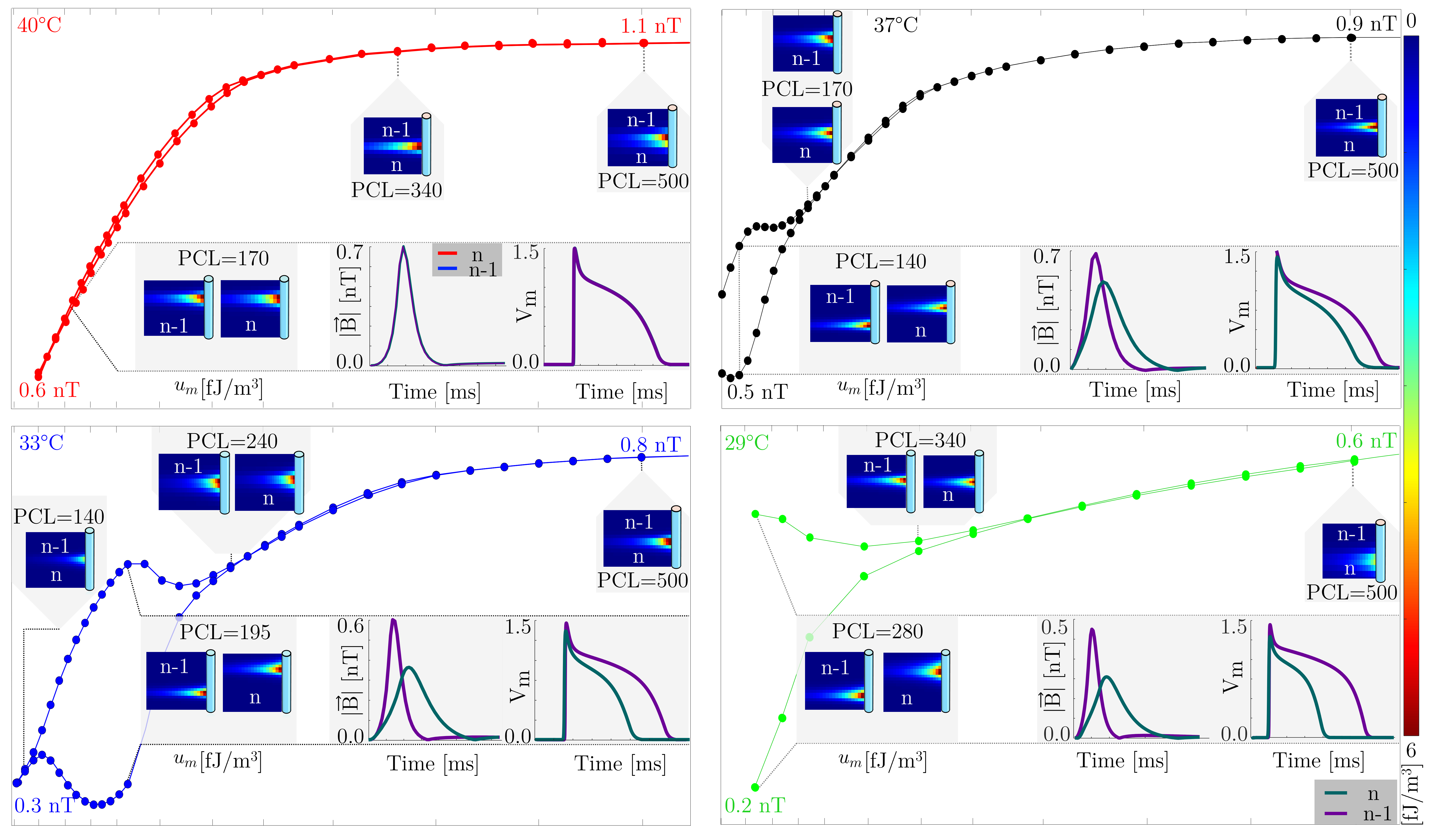}
   \caption{\small Representation of the magnetic energy distribution in the magnetic restitution curves. For each restitution curve we show the distribution along the wire of the magnetic energy density ($u_m$ [fJ/$m^3$]) for three selected PCLs, representative of steady-state, alternans onset and maximum alternans amplitude. Also, for the maximal alternans amplitude we show the action potentials and magnetic field norm for the $n$ (in  green) and $n-1$ (in purple) beat.}
    \label{fig:fig2}
\end{figure*}

\newpage

\section{Supplementary Methods}
\setcounter{equation}{0}  
\subsection{Cardiac Action Potential Modelling}
\renewcommand{\theequation}{S\arabic{equation}}
\label{cardicacmodel}
In this work, we simulate cardiac action potentials (AP) using a four variables phenomenological model including temperature-related effects on the cardiac electrophysiology~\cite{fenton2013role}.
Although more accurate models of cardiac action potentials exist, we adopted this model because of its capability to describe the AP features while maintaining a minimal level of complexity. The model reproduces the cardiac action potential using for variables \textit{u}, \textit{v}, \textit{w}, and \textit{s}. The variable \textit{u} represents the normalized membrane potential, while \textit{v}, \textit{w}, and \textit{s} are used to regulate fast inward ($J_{fi}$), slow inward ($J_{si}$), and slow outward ($J_{so}$) currents. In the following, we report the full set of equations starting from the equations for gating variables. 

\small
\begin{eqnarray}
\label{eqs1}
    \partial_t u&=& D\nabla^2u-(J_{fi}+J_{si}+J_{so}), \\
    \label{eqs2}
    \partial_t v&=&\Phi_v(T)\left[(1-H(u-\theta_v)) \frac{v_\infty -v}{\tau_v^-}-\frac{H(u-\theta_v)}{\tau_v^+} \right],\\
    \label{eqs3}
    \partial_t w&=&\Phi_w(T)\left[(1-H(u-\theta_w)) \frac{w_\infty -w}{\tau_w^-}-\frac{H(u-\theta_w)}{\tau_w^+} \right],\\
    \label{eqs4}
    \partial_t s&=& \Phi_s(T) \left[ \frac{(1+\tanh(k_s(u-u_s)))/2-s}{\tau_s} \right].
\end{eqnarray}
\normalsize

In Eq.~\ref{eqs1}, \textit{u} represents the normalized membrane potential, $D$ is the diffusion coefficient, and $J_{fi}$, $J_{si}$, $J_{so}$ are the fast-inward, slow-inward, and slow-outward currents, respectively. The three currents are voltage and temperature-dependent as explained by the following equations: 
\small
\begin{eqnarray}
    \label{eqs5}
    J_{fi}&=&\eta_{fi}(T)\left[H(u-\theta_v)(u-\theta_v)(u-u_u)\frac{v}{\tau_{fi}}\right],\\
    \label{eqs6}
    J_{si}&=&\eta_{si}(T)\left[H(u-\theta_w)\frac{ws}{\tau_{si}}\right],\\
    \label{eqs7}
    J_{so}&=&\eta_{so}(T)\left[ (1-H(u-\theta_w))\frac{(u_o-u)}{\tau_{o}}+\frac{H(u-\theta_w)}{\tau_{so}}\right].
\end{eqnarray}
\normalsize
In Eqs.~\ref{eqs1}-\ref{eqs7} the time constants are voltage-dependent, as detailed in the following: 
\small
\begin{eqnarray}
\label{eqs8}
\tau_v^-&=&(1-H(u-\theta_v^-))\tau_{v1}^{-}+-H(u-\theta_v^-)\tau_{v2}^{-},\\
\label{eqs9}
\tau_w^+&=& \tau_{w1}^+ +(\tau_{w2}^+ - \tau_{w1}^+) \frac{\tanh(k_w^+(u-u_w^+))+1}{2},\\
\label{eqs10}
\tau_w^-&=& \tau_{w1}^- +(\tau_{w2}^- - \tau_{w1}^-) \frac{\tanh(k_w^-(u-u_w^-))+1}{2},\\
\label{eqs11}
\tau_{so}^-&=& \tau_{so1} +(\tau_{so2} - \tau_{so1}) \frac{\tanh(k_{so}(u-u_{so}))+1}{2},\\
\label{eqs12}
\tau_s&=&(1-H(u-\theta_w))\tau_{s1}+-H(u-\theta_w)\tau_{s2},\\
\label{eqs13}
\tau_o&=&(1-H(u-\theta_o))\tau_{o1}+-H(u-\theta_o)\tau_{o2}.
\end{eqnarray}
\normalsize

In all the equations (Eqs.~\ref{eqs1}-\ref{eqs10}) above, $H(x)$ is the Heaviside step function, while $v$, $w$, and $s$ are three local gating variables. In Eqs.~\ref{eqs2}-\ref{eqs3}, $v_\infty$, and $w_\infty$ are defined by the following equations:
\small
\begin{eqnarray}
    \label{eqs14}
    v_\infty&=&H(\theta_v^- -u), \\
    \label{eqs15}
    w_\infty&=&(1-H(u-\theta_o)) \left( 1-\frac{u}{\tau_{w_\infty}}\right)+H(u-\theta_o)w_\infty^\star.
\end{eqnarray}
\normalsize

The effect of temperature on the action potential is described with two factors, $\Phi(T)$ and $\eta(T)$, affecting the kinetics of the gating variables and time constants of the ionic currents, respectively: 

\small
\begin{eqnarray}
    \label{eqs16}
    \Phi(T)&=&Q_{10}^{(T-T_A)/10},\\
    \label{eqs17}
    \eta(T)&=&A(1+B(T-T_A)).
\end{eqnarray}
\normalsize

where $T_A$ is the reference temperature of the tissue, while $A$, $B$ and $Q_{10}$ are constant coefficients obtained from experimental observations~\cite{kiyosue1993ionic,crozier1926distribution}. 
Model parameters are base on experimental optical mapping data~\cite{Crispino2024cross} and listed in Table~\ref{tab:table1}. 

\begin{table}[ht]
    \centering
    \begin{tabular}{|c|c||c|c||c|c|}
    \hline
    Parameter & Value & Parameter & Value & Parameter & Value \\ \hline
    $\theta_v$ & 0.3 & $\tau_{v^-_1}$ & 55 & $\tau_{w_1^+}$ & 175 \\ 
    $\theta_w$ & 0.13 & $\tau_{v^-_2}$ & 40 & $\tau_{w_2^+}$ & 230 \\ 
    $\theta_v^-$ & 0.2 & $\tau_v^+$ & 1.4506 & $\tau_{so_1}$ & 40 \\ 
    $\theta_o$ & 0.006 & $\tau_{w^-_1}$ & 40 & $\tau_{so_2}$ & 1.2 \\ 
    $\tau_{fi}$ & 0.10 & $\tau_{o1}$ & 470 & $k_{so}$ & 2 \\ 
    $\tau_{o2}$ & 6 & $u_w^+$ & 0.0005 & $u_{so}$ & 0.65 \\ 
    $k_w^-$ & 20 & $\tau_{s1}$ & 2.7342 & $\tau_{s2}$ & 2 \\ 
    $k_w^+$ & 8 & $k_{s}$ & 2.0994 & $u_{s}$ & 0.9087 \\ 
    $u_w^-$ & 0.00615 & $\tau_{si}$ & 2.9013 & $\tau_{w_\infty}$ & 0.0273 \\ 
    $w_\infty^\star$ & 0.78 & $Q_{10,v}$& 1.5 & $Q_{10,w}$ & 2.45 \\ 
    $ Q_{10,w}$ & 1.5 & $A_{fi}$& 1 & $B_{fi}$ & 0.065\\ 
    $A_{so}$ & 1 & $B_{so}$& 0.008 & $A_{si}$ & 1\\ 
    $B_{si}$ & 0.008 & & & &\\
    \hline
    \end{tabular}
    \caption{Model parameters for endocardial action potential minimal-model formulation for a selected thermal range (29$\div$40$^\circ$C). Units are given in ms, cm, mV, mS, $\mu$F, g, $^\circ$C.}
    \label{tab:table1}
\end{table}

The electrical activity is simulated in a 1D geometry, consisting in a 3~cm long cable (Figure \ref{fig:electrphysiology}).
We stimulated the wire with a full pacing down restitution protocol, which consists in delivering current pulses while decreasing the interval between two stimuli, i.e. the pacing cycle length (PCL). For each PCL we deliver 10 stimuli to reach the steady state, before decreasing the PCL. We started the stimulation protocol with a PCL of 800 ms and decreased it down to arrhythmic intervals until conduction block arises. We simulate the aforementioned dynamics for four different temperature, ranging from hyperthermia (40~$^\circ$C) to hypothermia (29~$^\circ$C) with a step of 4~$^\circ$C.
\renewcommand\thefigure{S\arabic{figure}} 
\begin{figure}
\setcounter{figure}{0}  
  \centering    \includegraphics[width=\textwidth]{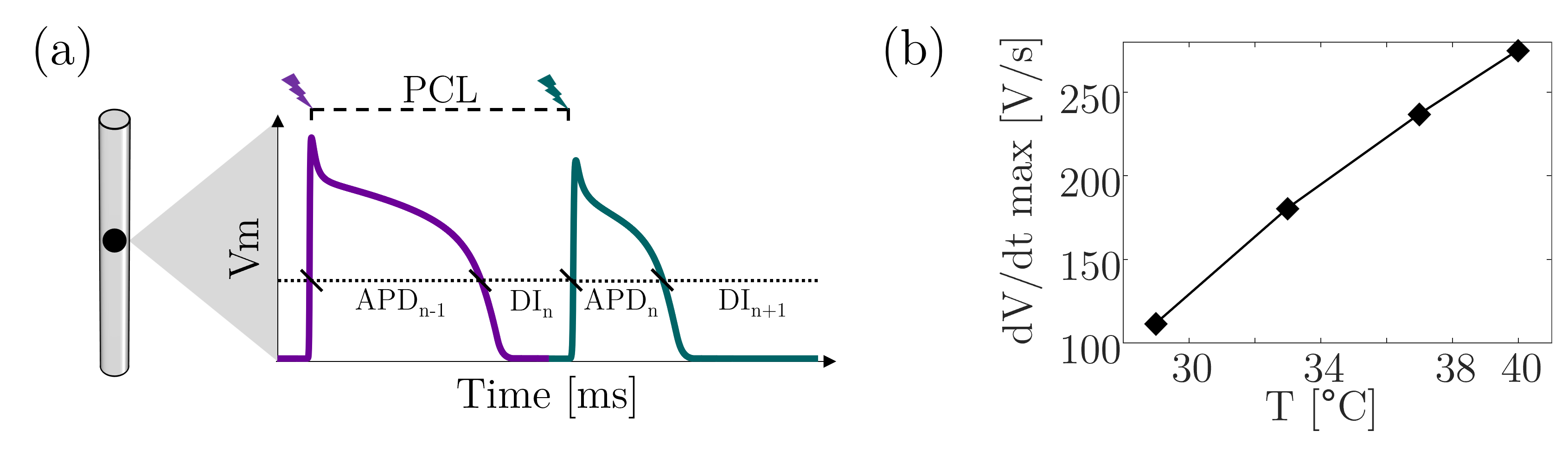}    \caption{(a) Schematic representation of two consecutive beats (n-1 and n) for a selected point along the wire with a graphical definition of action potential duration (APD), diastolic interval (DI), and pacing cycle  length (PCL). (b) Temperature-dependent behaviour of maximum derivative value of the depolarization phase of the AP for a selected beat.}   
\label{fig:electrphysiology} 
\end{figure}

\subsection{Alternans and conduction velocity measures}
It is well established that temperature affects cardiac dynamics by modifying several action potential features. In this study, we quantified APD$_{80}$ (Figure~\ref{fig:electrphysiology}) and conduction velocity, building restitution curves averaged over a previously detected time window for each PCL. 
\subsection{Magnetic Field Modelling}
In this work we calculate the magnetic field by applying the Bio-Savart law, as described in the main text. Briefly, we estimate the current density, $J(t,z')$, due to the action potential propagation in the \textit{z}-direction as follows:



\begin{equation}
\label{eqs19}
    J(t,z')=-\sigma \nabla V_m(t,z),
\end{equation}

where $\sigma$ represents the conductivity of the tissue, and $V_{m}(t,z)$ is the transmembrane potential expressed in mV. The numerical value of the conductivity is obtained applying the relation described in~\cite{fenton1998vortex} to the case of an homogeneous and isotropic fiber: 

\begin{equation}
    \label{eqs20}
    \sigma=D S_0 C_m,
\end{equation}

where $D=0.005$~cm/ms is the diffusion coefficient, $C_m=1~\mu$F/cm\textsuperscript{2} is the membrane capacitance, and $S_0$ is the surface to volume ratio calculated based on the standard dimensions of cardiac fibers, assimilated to a cylindrical compartment with a radius $R'=40\mu$~m, and a length $L=100~\mu$m. 
The transmembrane potential $V_{m}(t,z)$ is obtained from its dimensionless equivalent, $u$ (Eq.~\ref{eqs1}), applying the map $V_m=(83.3u-84)$~mV~\cite{fenton2013role}. 
The total current is computed considering the current density distributed in uniformly distributed in a small circular section $S'$, so that:

\begin{equation}
    I(t,z')=\int_{S'}\vec J(t,\vec r')\cdot d\vec S'= 
J(t,z')\pi R'^2
\end{equation}

This reasonable assumption is made to keep the computational complexity as low as possible, while effectively describing the physics of the problem. Indeed, this assumption allowed to compute the magnetic field by solving the Biot-Savart's integral considering a 1D straight line current source along the z-axis: 

\begin{equation}
\vec B_0=\frac{\mu_0}{4\pi}\int_{z_{min}}^{z_{max}}\frac{I(t,z')d\vec l' \wedge \Delta \vec r}{\vert \Delta \vec r\vert^3}
\end{equation}
with $\Delta \vec r=\vec r -\vec r'$, $\vec r =(x,y,z)$, $\vec r' =(0,0,z')$ and $\vec dl' =(0,0,dz')$. The \textit{x} and \textit{y} components of the field are then obtained by solving the following integrals: 

\begin{eqnarray}
\vec B_0&=&-\frac{\mu_0}{4\pi}\int_{z_{min}}^{z_{max}}\frac{I(t,z')y\,dz'}{[x^2+y^2+(z-z')^2]^{\frac32}}\hat i+\nonumber\\ &+&\frac{\mu_0}{4\pi}\int_{z_{min}}^{z_{max}}\frac{I(t,z')x\,dz'}{[x^2+y^2+(z-z')^2]^{\frac32}}\hat j\,.
\label{eq:magnetic_field}
\end{eqnarray}

\subsection{Numerical Methods}
The simulation of cardiac electrical activity and the calculation of the magnetic field have been performed in MATLAB R2022b (MathWorks, Natick, Massachusetts). The minimal model equations described in section~\ref{cardicacmodel} have been implemented using the finite-difference method to approximate both the spatial and temporal derivatives and solved using the Euler scheme~\cite{teukolsky1992numerical}. The final geometry consisted in a 3 cm long wire discretized with a mesh consisting of 300 nodes. We selected a suitable temporal discretization to ensure an accurate description of the AP and therefore of the temporal characteristics of the currents. 
The magnetic field is obtained solving the integral in Eq.~\ref{eq:magnetic_field} approximated with the trapezoidal method. To obtain an accurate description of the spatial behaviour of the magnetic field, we defined a spatial grid with a $10~\mu$m spacing in the \textit{x}, \textit{y}, and \textit{z} directions.

\end{document}